\documentclass[12pt]{article}
\usepackage{amssymb}
\usepackage{amsmath}
\usepackage{graphicx}
\usepackage{subfigure}
\usepackage{hyperref}

\begin{document}

\author{Ernst Trojan \and \textit{Moscow Institute of Physics and Technology} \and 
\textit{PO Box 3, Moscow, 125080, Russia}}
\title{Kink in superconducting cosmic string: exact solution}
\maketitle

\begin{abstract}
We solve the equations of motion and find the Lorentz transformation
associated with a kink in superconducting cosmic string. The kink velocity
does not depend on its amplitude. The kink amplitude cannot be arbitrary but
it varies within definite range and determines the explicit form of the
relevant Lorentz transformation.
\end{abstract}

\section{ Introduction}

The behavior of current-carrying (or superconducting) cosmic string is
determined by a pair of intrinsic equations of motion and a pair of
extrinsic equations where intrinsic and extrinsic motions are considered
separately \cite{C89a}. These equations can develop three classes of
perturbations \cite{C89a,C1,C2,MP00,CCMP02,C3,C4}, depending on the
parameter which is subject to change: shocks (the current), kinks (the
curvature), cusps (the spatial geometry). The shocks and kinks of
infinitesimal amplitude were studied analytically \cite{C89a}. Numerical
simulations \cite{MP00,CCMP02} revealed the process of finite-amplitude
shock and kink formation.

The numerical analysis becomes impossible when a stationary discontinuous
solution is formed. The analytical analysis remains the only helpful tool,
and it was already extended to the shocks with finite increment of the
current $\Delta \chi $ \cite{TV2012}. The kinks of finite amplitude (jump of
the curvature $\Delta \kappa ^\mu $) is the subject of our present interest.
We have recently adjusted the general theory of discontinuities \cite
{LL87,A89} to the superconducting cosmic strings and derived the equations
of intrinsic and extrinsic discontinuities \cite{T2013}. Now, we are looking
for explicit solution of these equations in application to the kinks.

\section{Characteristic four-vector}

The front of a discontinuity is a hypersurface , moving along the
characteristic direction in (3+1) space-time \cite{A89}. The discontinuity
is stable when its unit characteristic four-vector remains the same before
(label ''$-$'') and behind (label ''$+$'') the front \cite{T2013}: 
\begin{equation}
\lambda _{+}^\mu =\lambda _{-}^\mu \equiv \alpha _{+}\frac{w_{+}u_{+}^\mu
+v_{+}^\mu }{\sqrt{1-w_{+}^2}}+\sigma _{+}s_{+}^\mu =\alpha _{-}\frac{%
w_{-}u_{-}^\mu +v_{-}^\mu }{\sqrt{1-w_{-}^2}}+\sigma _{-}s_{-}^\mu
\label{lam11}
\end{equation}
\begin{equation}
\lambda _{\pm }^\mu \lambda _{\pm \mu }=1\qquad \alpha _{+}^2+\sigma
_{+}^2=\alpha _{-}^2+\sigma _{-}^2=1  \label{sta11}
\end{equation}
where unit worldsheet four-vectors $u_{\pm }^\mu $ ($u_{\pm }^\mu u_{\pm \mu
}=-1$), $v_{\pm }^\mu $ ($v_{\pm }^\mu v_{\pm \mu }=1$) and normal
four-vector $s_{\pm }^\mu $ ($s_{\pm }^\mu s_{\pm \mu }=1$) are mutual
orthogonal ($u_{\pm }^\mu v_{\pm \mu }=u_{\pm }^\mu s_{\pm \mu }=s_{\pm
}^\mu v_{\pm \mu }=0$). The velocities before and behind the front ($w_{-}$\
and $w_{+}$) are measured in the preferred reference frame, co-moving the
discontinuity. When we operate in the laboratory reference frame, the front
propagates at velocity $W$ with respect to the string, and there is no
motion before the front $W_{-}=0$ (however, the motion behind it $W_{+}\neq
0 $ may occur). When we switch to the co-moving reference frame, associated
with the front at rest, we operate with a finite flow before the front $%
w_{-}=-W$, and a flow behind the front is determined as $w_{+}=\left(
W_{+}-W\right) /\left( 1-Ww_{+}\right) $. 

There also exists a space-like unit four-vector $b_{\pm }^\nu $, which is
orthogonal to the relevant three four-vectors in (\ref{lam11}), namely:\ 
\begin{equation}
b_{\pm }^\nu u_{\pm \nu }=b_{\pm }^\nu v_{\pm \nu }=b_{\pm }^\nu s_{\pm \nu
}=0\qquad b_{\pm }^\nu b_{\pm \nu }=1  \label{bs}
\end{equation}
so that the sets $\left\{ u_{-}^\nu ,v_{-}^\nu ,s_{-}^\nu ,b_{-}^\nu
\right\} $ and $\left\{ u_{+}^\nu ,v_{+}^\nu ,s_{+}^\nu ,b_{+}^\nu \right\} $
constitute two orthonormal tetrads before and behind the discontinuity, and
the relevant Gram matrices $\gimel =diag\left( -1,1,1,1\right) $ coincide
with the Minkowski tensor. The Lorentz transformations

\begin{equation}
\left( 
\begin{array}{c}
u_{+}^\mu \\ 
v_{+}^\mu \\ 
s_{+}^\mu \\ 
b_{+}^\mu
\end{array}
\right) =O\left( 
\begin{array}{c}
u_{-}^\mu \\ 
v_{-}^\mu \\ 
s_{-}^\mu \\ 
b_{-}^\mu
\end{array}
\right) \qquad \left( 
\begin{array}{c}
u_{-}^\mu \\ 
v_{-}^\mu \\ 
s_{-}^\mu \\ 
b_{-}^\mu
\end{array}
\right) =O^{-1}\left( 
\begin{array}{c}
u_{+}^\mu \\ 
v_{+}^\mu \\ 
s_{+}^\mu \\ 
b_{+}^\mu
\end{array}
\right)  \label{lor}
\end{equation}
have unit determinant 
\begin{equation}
\det O=\det O^{-1}=\pm 1  \label{det1}
\end{equation}
and satisfy the orthogonality condition $O^T\gimel O=\gimel $ that is
equivalent to 
\begin{equation}
OO^{-1}=E=1\qquad O^{-1}O=\tilde E=1  \label{ort}
\end{equation}
where $E=\tilde E=diag\left( 1,1,1,1\right) $, and matrices 
\begin{equation}
O=\left( 
\begin{array}{cccc}
-Y & G & P & F \\ 
-H & X & Q & V \\ 
-A & R & S & I \\ 
-M & N & J & B
\end{array}
\right) \qquad O^{-1}=\left( 
\begin{array}{cccc}
-Y & H & A & M \\ 
-G & X & R & N \\ 
-P & Q & S & J \\ 
-F & V & I & B
\end{array}
\right)  \label{matr}
\end{equation}
include coefficients 
\begin{equation}
G=v_{-}^\mu u_{+\mu }\qquad H=u_{-}^\mu v_{+\mu }\qquad X=v_{-}^\mu v_{+\mu
}\qquad Y=u_{-}^\mu u_{+\mu }  \label{gh}
\end{equation}
\begin{equation}
P=s_{-}^\mu u_{+\mu }\qquad Q=s_{-}^\mu v_{+\mu }\qquad A=s_{+}^\mu u_{-\mu
}\qquad R=s_{+}^\mu v_{-\mu }  \label{pq}
\end{equation}
\begin{equation}
F=b_{-}^\mu u_{+\mu }\qquad V=b_{-}^\mu v_{+\mu }\qquad M=b_{+}^\mu u_{-\mu
}\qquad N=b_{+}^\mu v_{-\mu }  \label{mn}
\end{equation}
\begin{equation}
I=b_{-}^\mu s_{+\mu }\qquad J=b_{+}^\mu s_{-\mu }\qquad S=s_{-}^\mu s_{+\mu
}\qquad B=b_{+}^\mu b_{-\mu }  \label{ij}
\end{equation}
Identical transformation is achieved by (\ref{lor}) in the infinitesimal
limit 
\begin{equation}
X\rightarrow -Y\rightarrow S\rightarrow B\rightarrow 1\qquad G\rightarrow 0
\label{inf1}
\end{equation}
\begin{equation}
P\rightarrow Q\rightarrow F\rightarrow V\rightarrow I\rightarrow 0
\label{inf2}
\end{equation}
\begin{equation}
A\rightarrow R\rightarrow M\rightarrow N\rightarrow J\rightarrow 0
\label{inf3}
\end{equation}
when $w_{+}\rightarrow w_{-}$ and 
\begin{equation}
\alpha _{+}\rightarrow \alpha _{-}\qquad \sigma _{+}\rightarrow \sigma _{-}
\label{inf4}
\end{equation}

Multiplying equality (\ref{lam11}) by each of four-vectors $u_{-}^\mu $, $%
v_{-}^\mu $, $s_{-}^\mu $, $b_{-}^\mu $, we get a set of equations for the
characteristic four-vector:

\begin{equation}
\left( Yw_{+}+H\right) \frac{\alpha _{+}}{\sqrt{1-w_{+}^2}}+A\sigma _{+}=-%
\frac{w_{-}\alpha _{-}}{\sqrt{1-w_{-}^2}}  \label{ch1}
\end{equation}
\begin{equation}
\left( Gw_{+}+X\right) \frac{\alpha _{+}}{\sqrt{1-w_{+}^2}}+R\sigma _{+}=%
\frac{\alpha _{-}}{\sqrt{1-w_{-}^2}}  \label{ch2}
\end{equation}
\begin{equation}
\left( Pw_{+}+Q\right) \frac{\alpha _{+}}{\sqrt{1-w_{+}^2}}+S\sigma
_{+}=\sigma _{-}  \label{ch3}
\end{equation}
\begin{equation}
\left( Fw_{+}+V\right) \frac{\alpha _{+}}{\sqrt{1-w_{+}^2}}+I\sigma _{+}=0
\label{ch4}
\end{equation}
Multiplying equality (\ref{lam11}) by four-vectors $u_{+}^\nu $, $v_{+}^\nu $%
, $s_{+}^\nu $, $b_{+}^\nu $, we also get 
\begin{equation}
\left( Yw_{-}+G\right) \frac{\alpha _{-}}{\sqrt{1-w_{-}^2}}+P\sigma _{-}=-%
\frac{w_{+}\alpha _{+}}{\sqrt{1-w_{+}^2}}  \label{ch5}
\end{equation}
\begin{equation}
\left( Hw_{-}+X\right) \frac{\alpha _{-}}{\sqrt{1-w_{-}^2}}+Q\sigma _{-}=%
\frac{\alpha _{+}}{\sqrt{1-w_{+}^2}}  \label{ch6}
\end{equation}
\begin{equation}
\left( Aw_{-}+R\right) \frac{\alpha _{-}}{\sqrt{1-w_{-}^2}}+S\sigma
_{-}=\sigma _{+}  \label{ch7}
\end{equation}
\begin{equation}
\left( Mw_{-}+N\right) \frac{\alpha _{-}}{\sqrt{1-w_{-}^2}}+J\sigma _{-}=0
\label{ch8}
\end{equation}

The curvature four-vector \cite{C89b} 
\begin{equation}
\kappa _{\pm }^\mu =\eta _{+\rho }^\nu \nabla _\nu \eta _{+}^{\mu \rho }
\label{crv}
\end{equation}
is determined by the projective tensor 
\begin{equation}
\eta _{+}^{\mu \rho }=v_{\pm }^\mu v_{\pm }^\rho -u_{\pm }^\mu u_{\pm }^\rho
\label{eta1}
\end{equation}
In the light of (\ref{lor}) and (\ref{matr}), the latter is presented in the
form 
\begin{eqnarray}
\eta _{+}^{\mu \rho }=\left( H^2-Y^2\right) u_{-}^\mu u_{-}^\rho +\left(
GY-XH\right) u_{-}^{(\mu }v_{-}^{\rho )}+\left( X^2-G^2\right) v_{-}^\mu
v_{-}^\rho + &&  \nonumber \\
+\left( PY-QH\right) s_{-}^{(\mu }u_{-}^{\rho )}+\left( QX-PG\right)
s_{-}^{(\mu }v_{-}^{\rho )}+\left( Q^2-P^2\right) s_{-}^\mu s_{-}^\rho + && 
\nonumber \\
+\left( FY-VH\right) b_{-}^{(\mu }u_{-}^{\rho )}+\left( VX-FG\right)
b_{-}^{(\mu }v_{-}^{\rho )}+ &&  \nonumber \\
+\left( QV-PF\right) b_{-}^{(\mu }s_{-}^{\rho )}+\left( V^2-F^2\right)
b_{-}^\mu b_{-}^\rho \qquad &&  \label{eta2}
\end{eqnarray}
A finite-amplitude kink is associated with a finite change of the curvature $%
\kappa _{+}^\mu -\kappa _{-}^\mu \neq 0$, and the projective tensor (\ref
{eta1}) and (\ref{eta2}) cannot coincide at different sides of the
discontinuity. As soon as we know the components of matrix (\ref{matr}), we
know everything about the kink.

\section{Kink solution of extrinsic equations}

While the intrinsic equations of motion may admit discontinuities with
alternating current (shocks), the extrinsic equations may admit
discontinuity with alternating curvature (kinks) \cite{MP00,CCMP02}. Every
discontinuous solution of the extrinsic equations of motion satisfies the
following equations \cite{T2013}: 
\begin{equation}
\alpha _{+}\left\{ u_{-}^\mu -Hv_{+}^\mu +Yu_{+}^\mu +w_{+}\left( v_{-}^\mu
-Xv_{+}^\mu +Gu_{+}^\mu \right) \right\} =0  \label{lu11}
\end{equation}
\begin{equation}
\alpha _{-}\left\{ u_{+}^\mu -Gv_{-}^\mu +Yu_{-}^\mu +w_{-}\left( v_{+}^\mu
-Xv_{-}^\mu +Hu_{-}^\mu \right) \right\} =0  \label{lu22}
\end{equation}
\begin{equation}
\alpha _{+}\left\{ -w_{+}\left( u_{-}^\mu -Hv_{+}^\mu +Yu_{+}^\mu \right)
-c_{E+}^2\left( v_{-}^\mu -Xv_{+}^\mu +Gu_{+}^\mu \right) \right\} =0
\label{lu33}
\end{equation}
\begin{equation}
\alpha _{-}\left\{ -w_{-}\left( u_{+}^\mu -Gv_{-}^\mu +Yu_{-}^\mu \right)
-c_{E-}^2\left( v_{+}^\mu -Xv_{-}^\mu +Hu_{-}^\mu \right) \right\} =0
\label{lu44}
\end{equation}
where $c_E\equiv c_{E-}=c_{E+}$ is the speed of infinitesimal extrinsic
perturbations, and this speed is determined by the physical state of the
string \cite{C89a}. Since a kink solution admits the change of the
curvature, while the current $\chi _{+}=\chi _{-}$ remains constant, this
speed $c_E$ is the same at both sides.

Equations (\ref{lu11})-(\ref{lu44}) have evident degenerate solution 
\begin{equation}
\alpha _{+}=\alpha _{-}=0  \label{deg}
\end{equation}
The development of this degenerate discontinuity is not discussed now, it
will be considered in a separate study devoted to the cusps.\textrm{\ }As
for the non-degenerate kink discontinuities, functions $\alpha _{+}$ or $%
\alpha _{-}$ are not forbidden to vanish at some finite amplitude of the
perturbation rather than in the infinitesimal limit (\ref{inf4}).

If $\alpha _{+}\neq 0$ equations (\ref{lu11}) and (\ref{lu33}) have special
solution 
\begin{equation}
v_{-}^\mu =Xv_{+}^\mu -Gu_{+}^\mu  \label{v01}
\end{equation}
\begin{equation}
u_{-}^\mu =Hv_{+}^\mu -Yu_{+}^\mu  \label{u01}
\end{equation}
that, in the light of transformations (\ref{lor})-(\ref{matr}) gives 
\begin{equation}
P=Q=F=V=0  \label{pf00}
\end{equation}
\begin{equation}
A=R=M=N=0  \label{ar00}
\end{equation}
\begin{equation}
Y=\pm X=\pm \sqrt{1+G^2}\qquad H=\pm G  \label{h12}
\end{equation}
The same solution (\ref{pf00})-(\ref{h12}) satisfies equations (\ref{lu22})
and (\ref{lu44}) if $\alpha _{-}\neq 0$. Substituting (\ref{pf00})-(\ref{h12}%
) in (\ref{eta2}), we establish $\eta _{+}^{\mu \rho }=\eta _{-}^{\mu \rho }$
that implies constant curvature $\kappa _{+}^\mu =\kappa _{-}^\mu $ (\ref
{crv}) at any parameter $G$. Solution (\ref{pf00})-(\ref{h12}) cannot belong
to a finite-amplitude kink.

If $\alpha _{+}\neq 0$ and solution (\ref{pf00})-(\ref{h12}) takes no place,
equations (\ref{lu11}) and (\ref{lu33}) imply 
\begin{equation}
w_{+}=c_E  \label{wc1}
\end{equation}
\begin{equation}
u_{-}^\mu -Hv_{+}^\mu +Yu_{+}^\mu +c_E\left( v_{-}^\mu -Xv_{+}^\mu
+Gu_{+}^\mu \right) =0  \label{ee11}
\end{equation}
Multiplying equation (\ref{ee11}) by $u_{-}^\mu $, $v_{-}^\mu $, $s_{-}^\mu $%
, $b_{-}^\mu $, $u_{+}^\mu $, $v_{+}^\mu $, $s_{+}^\mu $, and $b_{+}^\mu $,
we obtain (omiting relations which are repeated): 
\begin{equation}
-1-\left( H+Xc_E\right) H+\left( Y+Gc_E\right) Y=0  \label{yy1}
\end{equation}
\begin{equation}
c_E-\left( H+Xc_E\right) X+\left( Y+Gc_E\right) G=0  \label{yy2}
\end{equation}
\begin{equation}
\left( H+Xc_E\right) Q-\left( Y+Gc_E\right) P=0  \label{yy3}
\end{equation}
\begin{equation}
\left( H+Xc_E\right) V-\left( Y+Gc_E\right) F=0  \label{yy4}
\end{equation}
\begin{equation}
A=-Rc_E  \label{yy5}
\end{equation}
\begin{equation}
M=-Nc_E  \label{yy6}
\end{equation}
If $\alpha _{-}\neq 0$ and solution (\ref{pf00})-(\ref{h12}) takes no place,
equations (\ref{lu22}) and (\ref{lu44}) imply 
\begin{equation}
w_{-}=c_E  \label{wc2}
\end{equation}
\begin{equation}
u_{+}^\mu -Gv_{-}^\mu +Yu_{-}^\mu +c_E\left( v_{+}^\mu -Xv_{-}^\mu
+Hu_{-}^\mu \right) =0  \label{xx11}
\end{equation}
Multiplying equation (\ref{xx11}) by $u_{-}^\mu $, $v_{-}^\mu $, $s_{-}^\mu $%
, $b_{-}^\mu $, $u_{+}^\mu $, $v_{+}^\mu $, $s_{+}^\mu $, and $b_{+}^\mu $,
we obtain 
\begin{equation}
-1-\left( G+Xc_E\right) G+\left( Y+Hc_E\right) Y=0  \label{xx1}
\end{equation}
\begin{equation}
c_E-\left( G+Xc_E\right) X+\left( Y+Hc_E\right) H=0  \label{xx2}
\end{equation}
\begin{equation}
\left( G+Xc_E\right) R-\left( Y+Hc_E\right) A=0  \label{xx3}
\end{equation}
\begin{equation}
\left( G+Xc_E\right) N-\left( Y+Hc_E\right) M=0  \label{xx4}
\end{equation}
\begin{equation}
P=-Qc_E  \label{xx5}
\end{equation}
\begin{equation}
F=-Vc_E  \label{xx6}
\end{equation}

Suppose that altogether $\alpha _{+}\neq 0$ and $\alpha _{-}\neq 0$. If $%
Q\neq 0$, equations (\ref{yy3}) and (\ref{xx5}) yield 
\begin{equation}
H+Xc_E=-c_E\left( Y+Gc_E\right)  \label{pq1}
\end{equation}
If $Q=0$, equation (\ref{xx5}) gives $P=0$, and equations (\ref{ch5})-(\ref
{ch6}) yield the same constraint (\ref{pq1}). Identity $H+Xc_E=0$ is
impossible because it is not valid at the infinitesimal level $X\rightarrow
1 $ and $H\rightarrow 0$. Substituting (\ref{pq1}) in equations (\ref{yy1})-(%
\ref{yy2}), we obtain solution

\begin{equation}
\allowbreak Y=\pm 1-Gc_E\qquad X=\mp 1-\frac G{c_E}\qquad H=G  \label{ki00}
\end{equation}

If $\alpha _{+}\neq 0$ and $\alpha _{-}=0$, \textrm{equations (\ref{ch5})
and (\ref{ch6}) result in (\ref{xx5}). Substituting (\ref{xx5})\ in (\ref
{yy3}) we get (\ref{pq1}) that together with (\ref{yy1})-(\ref{yy2}) yields (%
\ref{ki00}). Equation (\ref{ch8}) implies }$J=0$\textrm{, and taking
components of identity (\ref{ort}), namely } 
\begin{equation}
E_{14}=\left( Yc_E+G\right) N+FB=0  \label{e014}
\end{equation}
\begin{equation}
E_{24}=\left( Gc_E+X\right) N+VB=0  \label{e024}
\end{equation}
\textrm{together with (\ref{ki00}), we have } 
\begin{equation}
\left( F+Vc_E\right) B=0  \label{e124}
\end{equation}
\textrm{Since }$B=0$\textrm{\ contradicts to infinitesimal limit }$%
B\rightarrow 1$\textrm{, we must put }$B\neq 0$\textrm{\ and obtain equality
(\ref{xx6}).}

\textrm{If }$\alpha _{-}\neq 0$\textrm{\ and }$\alpha _{+}=0$\textrm{,
equations (\ref{ch1}) and (\ref{ch2}) result in (\ref{yy5}). Substituting (%
\ref{yy5})\ in (\ref{xx3}) we get } 
\begin{equation}
G+Xc_E=-c_E\left( Y+Hc_E\right)  \label{ar1}
\end{equation}
\textrm{where identity }$G+Xc_E=0$\textrm{\ is impossible. Substituting (\ref
{ar1}) in equations (\ref{xx1})-(\ref{xx2}), we obtain the same solution (%
\ref{ki00}). Equation (\ref{ch4}) implies }$I=0$\textrm{, making use of
relations incorporated in identity (\ref{ort}), namely} 
\begin{equation}
\tilde E_{14}=\left( Yc_E+G\right) N+MB=0  \label{ee014}
\end{equation}
\begin{equation}
\tilde E_{24}=\left( Gc_E+X\right) N+NB=0  \label{ee024}
\end{equation}
together with (\ref{ki00}), we have 
\begin{equation}
\left( M+Nc_E\right) B=0  \label{ee124}
\end{equation}
Since $B=0$\ contradicts to infinitesimal limit $B\rightarrow 1$, we must
put $B\neq 0$ and obtain equality (\ref{yy6}).

Thus, solution (\ref{ki00}) ta\textrm{kes place together with (\ref{yy5})-(%
\ref{yy6}), (\ref{xx5})-(\ref{xx6}). }Solution (\ref{ki00}) at the lower
sign 
\begin{equation}
\allowbreak Y=-1-Gc_E\qquad X=1-\frac G{c_E}\qquad H=G  \label{ki1}
\end{equation}
corresponds to identity $u_{+}^\mu =u_{-}^\mu $ and $v_{+}^\mu =v_{-}^\mu $
in the infinitesimal limit (\ref{inf1})-(\ref{inf3}). Solution (\ref{ki00})
with the upper sign corresponds to $u_{+}^\mu =-u_{-}^\mu $ and $v_{+}^\mu
=-v_{-}^\mu $ and implies space-time reversal, which was not discovered in
the infinitesimal perturbations \cite{C89a}. Therefore, this solution is not
realized, and we have to deal with the only one (\ref{ki1}).

By means of relations (\ref{yy5})-(\ref{yy6}), (\ref{xx5})-(\ref{xx6}) we
present matrices (\ref{matr}) in the form 
\begin{equation}
O=\left( 
\begin{array}{cccc}
-Y & G & -Qc_E & -Vc_E \\ 
-G & X & Q & V \\ 
Rc_E & R & S & I \\ 
Nc_E & N & J & B
\end{array}
\right) \qquad O^{-1}=\left( 
\begin{array}{cccc}
-Y & G & -Rc_E & -Nc_E \\ 
-G & X & R & N \\ 
Qc_E & Q & S & J \\ 
Vc_E & V & I & B
\end{array}
\right)  \label{matr1}
\end{equation}
Substituting (\ref{matr1}) in (\ref{ort}) we have 
\begin{equation}
E_{11}=Y^2-G^2-\left( Q^2+V^2\right) c_E^2=1  \label{e11}
\end{equation}
\begin{equation}
E_{22}=X^2-G^2+Q^2+V^2=1  \label{e22}
\end{equation}
so that $Y^2\geq 1$ and $Y^2\geq X^2$. In the light of (\ref{ki1}), these
inequalities are satisfied when 
\begin{equation}
0\leq G\leq G_{\max }=\frac{2c_E}{1-c_E^2}\qquad -1\leq g\leq 1
\label{range}
\end{equation}
where we define 
\begin{equation}
g=Gc_E+X=1+Gc_E-\frac G{c_E}  \label{gg}
\end{equation}
Since $\allowbreak Y<-1$ at any non-negative $G$, the Lorentz transformation
(\ref{lor}) is always orthochronous. Median value 
\begin{equation}
G_0=\frac{c_E}{1-c_E^2}\qquad X_0=-\frac{c_E^2}{1-c_E^2}<0  \label{med}
\end{equation}
is equivalent to $g=0$. When the current tends to the chiral limit $\chi
\rightarrow 0$, the velocity of extrinsic perturbations tends to the speed
of light $c_E\rightarrow 1$ \cite{C89a,MP00,CCMP02} that corresponds to $%
G_0\rightarrow \infty $ and $G_{\max }\rightarrow \infty $.

\section{Parameters of kink}

Substituting (\ref{ki1}) and (\ref{matr1}) in (\ref{ort}) we obtain a system
of algebraic equations. Omitting the evident repeated expressions, we write
down the following equations

\begin{equation}
E_{12}=G\left( g+1\right) -c_E\left( Q^2+V^2\right)  \label{e12}
\end{equation}
\begin{equation}
E_{23}=gR+QS+VI=0  \label{e23}
\end{equation}
\begin{equation}
E_{24}=gN+QJ+VB=0  \label{e24}
\end{equation}
\begin{equation}
E_{33}=\left( 1-c_E^2\right) R^2+S^2+I^2=1  \label{e33}
\end{equation}
\begin{equation}
E_{34}=\left( 1-c_E^2\right) RN+SJ+IB=0  \label{e34}
\end{equation}
\begin{equation}
E_{44}=\left( 1-c_E^2\right) N^2+J^2+B^2=1  \label{e44}
\end{equation}
\begin{equation}
\tilde E_{12}=G\left( g+1\right) -c_E\left( R^2+N^2\right) =0  \label{ee12}
\end{equation}
\begin{equation}
\tilde E_{23}=gQ+RS+NJ=0  \label{ee23}
\end{equation}
\begin{equation}
\tilde E_{24}=gV+RI+NB=0  \label{ee24}
\end{equation}
\begin{equation}
\tilde E_{33}=\left( 1-c_E^2\right) Q^2+S^2+J^2=1  \label{ee33}
\end{equation}
\begin{equation}
\tilde E_{34}=\left( 1-c_E^2\right) QV+SI+JB=0  \label{ee34}
\end{equation}
\begin{equation}
\tilde E_{44}=\left( 1-c_E^2\right) V^2+I^2+B^2=1  \label{ee44}
\end{equation}
where $g$ is defined in (\ref{gg}).

The unitary determinant (\ref{det1}) of the Lorentz matrices (\ref{matr1})
is:

\begin{equation}
\left( \allowbreak G^2-YX\right) \left( BS-IJ\right) +\left\{ \left(
Y+c_EG\right) +c_E\left( c_EX+G\right) \right\} \left\{ R\left( QB-VJ\right)
+N\allowbreak \left( VS-QI\right) \right\} =\pm 1  \label{det2}
\end{equation}
Substituting (\ref{ki1}), (\ref{gg}) and (\ref{e12}), (\ref{e23}), (\ref{e24}%
) in (\ref{det2}), we have

\begin{equation}
BS-IJ=\pm g  \label{det3}
\end{equation}
that is equivalent to 
\begin{equation}
\left( S^2+J^2\right) \left( I^2+B^2\right) -\left( IS+JB\right) ^2=g^2
\label{det4}
\end{equation}
and contains no additional information because it is a consequence of
equations (\ref{e12}), (\ref{ee33}), (\ref{ee34}), (\ref{ee44}).

Taking into account solution (\ref{yy5})-(\ref{yy6}), (\ref{xx5})-(\ref{xx6}%
) and (\ref{ki1}), we simplify equations (\ref{ch1})-(\ref{ch8}) in the
following form 
\begin{equation}
g\alpha _{+}+\sqrt{1-c_E^2}R\sigma _{+}=\alpha _{-}  \label{ch22}
\end{equation}
\begin{equation}
\sqrt{1-c_E^2}Q\alpha _{+}+S\sigma _{+}=\sigma _{-}  \label{ch33}
\end{equation}
\begin{equation}
\sqrt{1-c_E^2}V\alpha _{+}+I\sigma _{+}=0  \label{ch44}
\end{equation}
\begin{equation}
g\alpha _{-}+\sqrt{1-c_E^2}Q\sigma _{-}=\alpha _{+}  \label{ch66}
\end{equation}
\begin{equation}
\sqrt{1-c_E^2}R\alpha _{-}+S\sigma _{-}=\sigma _{+}  \label{ch77}
\end{equation}
\begin{equation}
\sqrt{1-c_E^2}N\alpha _{-}+J\sigma _{-}=0  \label{ch88}
\end{equation}
Equations (\ref{ch22})-(\ref{ch88}) determine the unknowns $\alpha _{\pm }$
and $\sigma _{\pm }$ but they do not\ provide enough information to
equations (\ref{e12})-(\ref{ee44}).

It is easy to check that equation (\ref{e34}) is a consequence of (\ref{e23}%
), (\ref{e33}), (\ref{ee24}), (\ref{ee34}), (\ref{ee44}). In turn, equation (%
\ref{ee24}) is a consequence of (\ref{e12}), (\ref{e23}), (\ref{e24}), (\ref
{ee33}), (\ref{ee44}). Equation (\ref{e44}) follows from (\ref{e12}), (\ref
{e33}), (\ref{ee12}), (\ref{ee33}), (\ref{ee44}). But equation (\ref{ee12})
follows from (\ref{e23}), (\ref{e24}), (\ref{ee33}), (\ref{ee34}), (\ref
{ee44}). Equation (\ref{ee23}) is derived from (\ref{e12}), (\ref{e23}), (%
\ref{e24}), (\ref{ee33}), (\ref{ee34}). After all, taking equations (\ref
{e12}), (\ref{e23}), (\ref{ee33}), (\ref{ee33}), (\ref{ee44}), we come to
equation (\ref{ee34}). As a result, there are only 6 independent equations (%
\ref{e12}), (\ref{e23}), (\ref{e24}), (\ref{e33}), (\ref{ee33}), and (\ref
{ee44}), while there are 9 unknowns $Q$, $R$, $N$, $V$, $S$, $B$, $I$, $J$, $%
G$. These 6 independent equations allow to determine 6 variables, while the
rest 3 variables, for example, $Q$, $R$, $G$ cannot be reduced or expressed
through each other. As soon as we know the triplet $\left\{ Q,R,G\right\} $,
all other unknowns are immediately established.

Nevertheless, the system must be fully resolvable because we have started
with two extrinsic equations of motion and two independent parameters. The
first parameter is the kink velocity $c_E$, it is already determined by the
physical state of the string. The second parameter is an increment of the
curvature $\Delta \kappa ^\mu $, it can be reflected by a single variable.
Therefore, unknowns $\left\{ Q,R,G\right\} $ are no more than functions,
depending on this singular argument. The most natural choice concerns $G$,
and we must establish functions $Q\left( G\right) $ and $R\left( G\right) $.

\section{Energy of kink}

Defining 
\begin{equation}
Q=h\sin a\qquad V=h\cos a  \label{sinq}
\end{equation}
\begin{equation}
R=h\sin x\qquad N=h\cos x  \label{sinr}
\end{equation}
where 
\begin{equation}
h^2=\frac{G\left( g+1\right) }{c_E}=\frac{1-g^2}{1-c_E^2}=\frac{\left(
1-c_E^2\right) }{c_E^2}G\left( G_{\max }-G\right)  \label{hh2}
\end{equation}
and $G_{\max }$ is taken from (\ref{range}), we present equations (\ref{e12}%
) and (\ref{ee12}) in the form

\begin{equation}
Q^2+V^2=R^2+N^2=h^2  \label{hqr}
\end{equation}
Function $h\left( G\right) $ disappears when $G=0$ ($g=1$) or when $%
G=G_{\max }$ ($g=-1$) that takes place when and only when 
\begin{equation}
Q=V=0\qquad \Leftrightarrow \qquad R=N=0  \label{hqr0}
\end{equation}
In the light of (\ref{yy5})-(\ref{yy6}) and (\ref{xx5})-(\ref{xx6}), it
corresponds to (\ref{pf00})-(\ref{ar00}), while equalities (\ref{e11})-(\ref
{e22}) imply $Y^2=X^2=1+G^2$. As a result, the projective tensor (\ref{eta2}%
) remains unchanged and the curvature is constant when $h=0$.

A triplet of unknowns $\left\{ a,x,h\right\} $ is equivalent to $\left\{
Q,R,G\right\} $. Defining four-vectors 
\begin{equation}
h_{-}^\mu =\left( Qs_{-}^\mu +Vb_{-}^\mu \right) =h\left( s_{-}^\mu \sin
a+b_{-}^\mu \cos a\right)  \label{hqu1}
\end{equation}
\begin{equation}
h_{+}^\mu =\left( Rs_{+}^\mu +Nb_{+}^\mu \right) =h\left( s_{+}^\mu \sin
x+b_{+}^\mu \cos x\right)  \label{hqu2}
\end{equation}
\begin{equation}
\zeta _{\pm }^\mu =G\left( c_Eu_{\pm }^\mu +v_{\pm }^\mu \right) -c_Eh_{\pm
}^\mu  \label{zeta1}
\end{equation}
where 
\begin{equation}
\zeta _{\pm }^\mu \zeta _{\pm \mu }=2Gw\qquad h_{\pm }^\mu h_{\pm \mu }=h^2
\label{zh2}
\end{equation}
we substitute them in (\ref{lor}), (\ref{matr1}) and obtain relations 
\begin{equation}
\zeta _{+}^\mu =-\zeta _{-}^\mu  \label{z-z}
\end{equation}
\begin{equation}
u_{\pm }^\mu =u_{\mp }^\mu +\zeta _{\mp }^\mu \qquad v_{\pm }^\mu =v_{\mp
}^\mu -\frac{\zeta _{\mp }^\mu }{c_E}  \label{zuv}
\end{equation}
Substituting (\ref{zuv}) in (\ref{eta1}), we determine the projective tensor
behind the discontinuity 
\begin{equation}
\eta _{+}^{\mu \rho }=\left( v_{-}^\mu -\frac{\zeta _{-}^\mu }{c_E}\right)
\left( v_{-}^\rho -\frac{\zeta _{-}^\rho }{c_E}\right) -\left( u_{-}^\mu
+\zeta _{-}^\mu \right) \left( u_{-}^\rho +\zeta _{-}^\rho \right)
\label{eta3}
\end{equation}
that is 
\begin{equation}
\eta _{+}^{\mu \rho }=\eta _{-}^{\mu \rho }-\frac{v_{-}^{(\mu }\zeta
_{-}^{\rho )}}{c_E}-u_{-}^{(\mu }\zeta _{-}^{\rho )}+\frac{1-c_E^2}{c_E^2}%
\zeta _{-}^\mu \zeta _{-}^\rho  \label{eta4}
\end{equation}
In the light of (\ref{lor}) and (\ref{matr1}), we established an invariant
time-like bicharacteristic four-vector 
\begin{equation}
\xi ^\mu =\xi _{-}^\mu \equiv u_{-}^\mu +c_Ev_{-}^\mu =\xi _{+}^\mu \equiv
u_{+}^\mu +c_Ev_{+}^\mu  \label{xi1}
\end{equation}
which is orthogonal to the characteristic four-vector (\ref{lam11}) and to
four-vectors (\ref{hqu1})-(\ref{zeta1}).

The stress-energy tensor of a cosmic string \cite{C89b} 
\begin{equation}
\bar T_{\pm }^{\mu \nu }=Uu_{\pm }^\mu u_{\pm }^\nu -Tv_{\pm }^\mu v_{\pm
}^\nu =U\left( u_{\pm }^\mu u_{\pm }^\nu -c_E^2v_{\pm }^\mu v_{\pm }^\nu
\right)  \label{ten0}
\end{equation}
includes the same values $U_{+}=U_{-}\equiv U$ and $T_{+}=T_{-}\equiv T$
because the current $\chi $ is the same at both sides of the discontinuity.
Taking into account expressions (\ref{zuv}) and (\ref{xi1}), we have 
\begin{equation}
\bar T_{+}^{\mu \nu }=\bar T_{-}^{\mu \nu }+U\xi _{-}^{(\mu }\zeta _{-}^{\nu
)}  \label{ten1}
\end{equation}
The energy density 
\begin{equation}
E_{+}=\bar T_{+}^{00}=E_{-}+2U\xi _{-}^0\zeta _{-}^0=E_{-}-2U\xi _{+}^0\zeta
_{+}^0  \label{ten00}
\end{equation}
in the light of (\ref{hqu1})-(\ref{zeta1}), is written so: 
\begin{equation}
E_{+}=E_{-}+E_0\left( wu_{-}^0+v_{-}^0\right) G-E_0\left( s_{-}^0\sin
a+b_{-}^0\cos a\right) c_Eh\left( G\right)  \label{enf1}
\end{equation}
\begin{equation}
E_{+}=E_{-}-E_0\left( wu_{+}^0+v_{+}^0\right) G+E_0\left( s_{+}^0\sin
x+b_{+}^0\cos x\right) c_Eh\left( G\right)  \label{enf2}
\end{equation}
where 
\begin{equation}
E_0=2U\xi _{-}^0=2U\xi _{+}^0  \label{enf0}
\end{equation}
is independent on the kink parameters because it is determined by the string
worldsheet and the current which circulates within it.

We consider a finite-amplitude stationary kink with constant characteristic
four-vector (\ref{lam11}) rather than a kink in the process of formation and
development. As a matter of fact, as soon as a kink is formed, the energy is
not conserved \cite{MP00,CCMP02}. Although this stationary kink is
responsible for transition between two equilibrium states of the string, the
energy is not conserved ($E_{+}\neq E_{-}$) because it is not a continuous
solution of the equations of motion. The same variational principle cannot
be applied immediately at both sides of the discontinuity, but the energy
extremum takes place with respect to the relevant floating parameters at
each side. It implies the following.


The energy functional (\ref{enf1}) becomes a unique function $E_{+}\left\{
G\right\} $ when the floating variable $a$ is adjusted at the extremum level 
\begin{equation}
\frac{\partial E_{+}}{\partial a}=0\qquad \Rightarrow \qquad s_{-}^0\cos
a=b_{-}^0\sin a\qquad \Leftrightarrow \qquad s_{-}^0V=b_{-}^0Q  \label{extr1}
\end{equation}
In the view of (\ref{lor}) and (\ref{matr1}), equality (\ref{extr1}) implies

\begin{equation}
\left( NS-RJ\right) s_{-}^0=0\qquad \left( NI-RB\right) b_{-}^0=0
\label{sb11}
\end{equation}

There is no principal difference between the sides of the discontinuity: in
contrast to the shock waves \cite{TV2012}, the evolutionary condition
imposes no restriction on the discontinuity transition in both directions
(from ''$+$'' to ''$-$'' and the inverse direction). \textrm{However, we
cannot deal with fixed }$u_{-}^\mu $\textrm{, }$v_{-}^\mu $\textrm{, }$%
s_{-}^\mu $\textrm{, }$b_{-}^\mu $\textrm{\ and fixed }$u_{+}^\mu $\textrm{, 
}$v_{+}^\mu $\textrm{, }$s_{+}^\mu $\textrm{, }$b_{+}^\mu $\textrm{\
simultaneously, since only one set is given, while the other set is
determined by the kink, according to the Lorentz transformation} (\ref{lor}%
). As soon as we operate with given four-vectors $u_{+}^\mu $, $v_{+}^\mu $, 
$s_{+}^\mu $, $b_{+}^\mu $ as a primary tetrad, the energy functional (\ref
{enf2}) is considered as unique function $E_{+}\left\{ G\right\} $ where the
floating variable $x$ is adjusted at the extremum level\textrm{\ } 
\begin{equation}
\frac{\partial E_{+}}{\partial x}=0\qquad \Rightarrow \qquad s_{+}^0\cos
x=b_{+}^0\sin x\qquad \Leftrightarrow \qquad s_{+}^0N=b_{+}^0R  \label{extr2}
\end{equation}
In the view of \ref{lor}) and (\ref{matr1}), equality (\ref{extr2}) implies 
\begin{equation}
\left( VS-QJ\right) s_{+}^0=0\qquad \left( VI-QB\right) b_{+}^0=0
\label{sb22}
\end{equation}

By means of (\ref{lor}) and (\ref{matr1}), we establish the following
possibilities 
\begin{equation}
s_{-0}=0\qquad \vec \xi _{-}\cdot \vec s_{-}=w\left( \vec u_{+}^2+\vec
v_{+}^2\right) Q=0\quad \Rightarrow \quad Q=0  \label{sbb1}
\end{equation}
\begin{equation}
b_{-0}=0\qquad \vec \xi _{-}\cdot \vec b_{-}=w\left( \vec u_{+}^2+\vec
v_{+}^2\right) V=0\quad \Rightarrow \quad V=0  \label{sbb2}
\end{equation}
\begin{equation}
s_{+0}=0\qquad \vec \xi _{+}\cdot \vec s_{+}=w\left( \vec u_{+}^2+\vec
v_{+}^2\right) R=0\quad \Rightarrow \quad R=0  \label{sbb3}
\end{equation}
\begin{equation}
b_{+0}=0\qquad \vec \xi _{+}\cdot \vec b_{+}=w\left( \vec u_{+}^2+\vec
v_{+}^2\right) N=0\quad \Rightarrow \quad N=0  \label{sbb4}
\end{equation}
where vectors $\vec u_{\pm }$, $\vec v_{\pm }$, $\vec s_{\pm }$, $\vec
b_{\pm }$ and $\vec \xi _{\pm }=\vec u_{\pm }+c_E\vec v_{\pm }$ are spatial
components of the relevant four-vectors. In the view of (\ref{hqr0}) and (%
\ref{sbb1})-(\ref{sbb4}), it is clear that simultaneous $s_{-}^0=b_{-}^0=0$
or simultaneous $s_{+}^0=b_{+}^0=0$ corresponds to constant curvature at
zero kink amplitude $G=0$. Equations (\ref{sb11}) and (\ref{sb22}) are
satisfied without regard of $s_{\pm }^0$ and $b_{\pm }^0$ when, again,
identity (\ref{hqr0}) takes place and the curvature remains constant. After
all, simultaneous non-zero $s_{-}^0\neq 0$, $b_{-}^0\neq 0$ in (\ref{sb11})
or simultaneous non-zero $s_{+}^0\neq 0$, $b_{+}^0\neq 0$ in (\ref{sb22})
implies $BS-IJ=0$, in the light of (\ref{det3}), it is corresponds to
particular $g=0$ that does not embrace the infinitesimal limit at $%
g\rightarrow 1$ ($G\rightarrow 0$).

Therefore, a kink of finite amplitude can satisfy equations (\ref{sb11}) and
(\ref{sb22}) under one of the following conditions: 
\begin{equation}
s_{-0}=s_{+0}=0\qquad Q=R=0  \label{fol1}
\end{equation}
\begin{equation}
s_{-0}=b_{+0}=0\qquad Q=N=0  \label{fol2}
\end{equation}
\begin{equation}
b_{-0}=s_{+0}=0\qquad V=R=0  \label{fol3}
\end{equation}
\begin{equation}
b_{-0}=b_{+0}=0\qquad V=N=0  \label{fol4}
\end{equation}
As a matter of fact, when 8 components of four-vectors $s_{-}^\mu $ and $%
b_{-}^\mu $ are subject to 7 constraints, and 1 component remains uncertain,
the discontinuity is running in uncertain direction. The problem remains
disambiguous until we impose additional self-consistent constraint
associated with extremum of the energy density (\ref{enf1}) where only one
component ($s_{-}^0$ or $b_{-}^0$) is free to float, while the other
component is fixed at a value which provides the energy minimum. The same
view concerns the components of four-vectors $s_{+}^\mu $ and $b_{+}^\mu $.

\section{Explicit kink solution}

Substituting constraint (\ref{fol1}) in equations (\ref{ch22}) and (\ref
{ch66}), we find $g^2=1$ (because we do not consider the degenerate case $%
\alpha _{-}=0$) that, in the light of (\ref{hh2}), corresponds to constant
curvature. Substituting (\ref{fol2}) in equation (\ref{e24}), we find $VB=0$
where solution $B=0$ contradicts to the infinitesimal limit (\ref{inf1})
while solution $V=0$, in the view of (\ref{e12}) and (\ref{fol2}), implies
constant curvature. Substituting (\ref{fol3}) in equation (\ref{ee24}), we
find $NB=0$ where solution $B=0$ is impossible while solution $V=0$, in the
view of (\ref{ee12}) and (\ref{fol3}), again implies constant curvature.

Substituting constraint (\ref{fol4}) in equations (\ref{e12}) and (\ref{ee12}%
) we find 
\begin{equation}
R^2=Q^2=\frac{1-g^2}{1-c_E^2}  \label{rq1}
\end{equation}
where $Q=$ $R=0$ at constant curvature. Substituting constraint (\ref{fol4})
in equations (\ref{e24}) and (\ref{ee24}), we find 
\begin{equation}
I=J=0  \label{ij0}
\end{equation}
Substituting (\ref{ij}) in equation (\ref{e23}) and (\ref{e44}), we find $%
S=\pm g$ and $B^2=1$ that, in the view of infinitesimal limit (\ref{inf1})
at $g\rightarrow 1$ (\ref{gg}), implies 
\begin{equation}
S=g\qquad B=1  \label{bs1}
\end{equation}
Substituting (\ref{bs1}) back in (\ref{e23}), we find 
\begin{equation}
R=-Q=\pm \sqrt{G\left( 2-G\frac{\left( 1-c_E^2\right) }{c_E^2}\right) }
\label{rq11}
\end{equation}
Substituting solution (\ref{rq1})-(\ref{rq11}) in equations (\ref{ch22})-(%
\ref{ch88}), we have two independent equations 
\begin{equation}
\alpha _{+}=g\alpha _{-}+\sqrt{1-w^2}Q\sigma _{-}  \label{alf6}
\end{equation}
\begin{equation}
\sigma _{+}=-\sqrt{1-w^2}Q\alpha _{-}+g\sigma _{-}  \label{sig6}
\end{equation}
Solution (\ref{rq1})-(\ref{sig6}) describes a kink of finite amplitude.

Not\textrm{e that conditions of extremum (\ref{extr1}) and (\ref{extr2})
allow to present the energy density (\ref{enf1})-(\ref{enf2}) in the for}m
of functional 
\begin{equation}
E_{+}\left[ G\right] =E_{-}+E_0\left( c_Eu_{-}^0+v_{-}^0\right)
G-E_0c_Es_{-}^0Q  \label{enf11}
\end{equation}
where $Q$ () is a unique function on $G$. As we have mentioned above, a kink
can propagate in both directions that is reflected in the choice of the sign
in formula (\ref{rq11}). The energy density (\ref{enf1})-(\ref{enf2}) of
low-amplitude kinks at small $G$ and $s_{+}^0\simeq s_{-}^0=s^0$ is given by
expression 
\begin{equation}
E_{+}\simeq E_{-}-E_0c_Es^0Q\simeq E_{-}\mp E_0s^0\sqrt{2Gc_E}
\label{small1}
\end{equation}
Inequality $E_{+}<E_{-}$ implies an energetically favorable regime of
propagation from ''$-$'' to ''$+$''. Opposite inequality $E_{+}>E_{-}$
corresponds to the kink propagation from ''$+$'' to ''$-$''. The critical
kink development at $G=G_{\max }$ in (\ref{range}) takes place at $g=S=-1$
in (\ref{bs1}) and corresponds to constant curvature. The maximum absolute
amplitude 
\begin{equation}
R=-Q=\pm \frac 1{\sqrt{1-c_E^2}}\qquad  \label{rq111}
\end{equation}
is achieved at the median value $G=G_0$ (\ref{med}), corresponding to $g=S=0$%
. Indeed, function $R(g)=-Q(g)$ from (\ref{rq1}) is symmetric around its
maximum at $g=0$ and comes to minimum at $g=\pm 1$.

\section{Conclusion}

In contrast to the shocks, the kinks require much more complicated analysis.
The equations of extrinsic discontinuities (\ref{lu11})-(\ref{lu44}) give no
evident hint to explicit solution. We have to consider the Lorentz
transformation (\ref{lor}) between the orthonormal tetrads before and behind
the discontinuity, where the matrices of transformation (\ref{matr}) are not
arbitrary but depend on the kink amplitude. The latter is associated with
parameter $G$, which is not arbitrary but can vary within a definite finite
range (\ref{range}). According to the energy extremum, we establish the
elements of the transformation (\ref{matr1}) (\ref{ki1}), (\ref{rq1})-(\ref
{rq11}). The kink velocity coincides with the speed of infinitesimal
extrinsic perturbations $c_E$ (\ref{wc1}), (\ref{wc2}) and does not depend
on the kink amplitude $G$. The spatial geometry of the string is changing
within a 2-dimensional plane because four-vector $b_{\pm }^\mu $ remains
constant at both sides of the discontinuity.

Having derived the kink solution (\ref{lor}), (\ref{matr1}) (\ref{ki1}), (%
\ref{rq1})-(\ref{rq11}), we are ready to look for its further application by
means of equations (\ref{crv})-(\ref{eta2}), (\ref{alf6})-(\ref{sig6}). Now,
the following problem deserves consideration: an explicit link between the
kink parameter $G$, the relative increment of the curvature $\Delta \kappa
^\mu /\kappa _{-}^\mu $ and the angle in the kink vertex.


\begin{thebibliography}{99}
\bibitem{C89a}  B. Carter, Phys. Lett. B 228, 466 (1989).

\bibitem{C1}  A. Vilenkin and E. P. S. Shellard, Cosmic strings and other
topological defects, (Cambridge University Press, 2000) p. 159.

\bibitem{C2}  J. J. Blanco-Pillado and K. D. Olum, Phys. Rev. D 59, 063508
(1999). arXiv:gr-qc/9810005 

\bibitem{MP00}  X. Martin and P. Peter, Phys. Rev. D 61, 043510 (2000).
arXiv:hep-ph/9808222

\bibitem{CCMP02}  A. Cordero-Cid, X. Martin, and P. Peter, Phys.Rev. D 65,
083522 (2002). arXiv:hep-ph/0201097

\bibitem{C3}  E. J. Copeland and T. W. B. Kibble, Phys. Rev. D 80, 123523
(2009). arXiv:0909.1960 

\bibitem{C4}  C. J Copi and T. Vachaspati, Phys. Rev. D 83, 023529 (2011).
arXiv:1010.4030 


\bibitem{TV2012}  E. Trojan and G.V. Vlasov, Phys. Rev. D 85, 107303 (2012).
arXiv:1102.5659

\bibitem{LL87}  L.D. Landau and E.M. Lifshitz, \textit{Fluid mechanics}, 2nd
ed. (Pergamon, Oxford, 1987), p. 331.

\bibitem{A89}  A. M. Anile, \textit{Relativistic fluids and magneto-fluids},
(Cambridge University Press, 1989), p. 215.

\bibitem{T2013}  E. Trojan, arXiv::1312.4845

\bibitem{C89b}  B. Carter, Phys. Lett. B 224, 61 (1989).

\end{thebibliography}
\end{document}